\documentclass[prb,twocolumn,superscriptaddress,showpacs]{revtex4}
\usepackage{graphicx}
\usepackage{dcolumn}
\usepackage{bm}

\begin{document}

\title{NQR and X-ray investigation of the structure of Na$_{2/3}$CoO$_{2}$ compound}

\author{T.A.~Platova}
\affiliation{Physics Department, Kazan State University, 420008 Kazan, Russia}
\affiliation{Laboratoire de Physique des Solides, UMR 8502, Universit\'e
Paris-Sud, 91405 Orsay, France}
\author{I.R.~Mukhamedshin}
\email{Irek.Mukhamedshin@ksu.ru}%
\affiliation{Physics Department, Kazan State University, 420008 Kazan, Russia}
\affiliation{Laboratoire de Physique des Solides, UMR 8502, Universit\'e
Paris-Sud, 91405 Orsay, France}
\author{H.~Alloul}
\affiliation{Laboratoire de Physique des Solides, UMR 8502, Universit\'e
Paris-Sud, 91405 Orsay, France}
\author{A.V.~Dooglav}
\affiliation{Physics Department, Kazan State University, 420008 Kazan, Russia}
\affiliation{Laboratoire de Physique des Solides, UMR 8502, Universit\'e
Paris-Sud, 91405 Orsay, France}
\author{G.~Collin}
\affiliation{Laboratoire de Physique des Solides, UMR 8502, Universit\'e
Paris-Sud, 91405 Orsay, France}

\begin{abstract}
We have synthesized various samples of the $x=2/3$ phase of sodium cobaltate
Na$_{x}$CoO$_{2}$ and performed X-ray powder diffractions spectra to compare
the diffraction with the structure proposed previously from NMR/NQR
experiments [H.~Alloul \emph{et al.}, EPL \textbf{85}, 47006 (2009)]. Rietveld analysis of the data are found in perfect agreement with those, and confirm the concentration x=2/3 obtained in the synthesis procedure. They even give indications on the atomic displacements of Na inside the unit cell. The detailed NQR data allow us to identify the NQR transitions and electric field gradient (EFG) parameters for 4 cobalt sites and 3 Na sites. The spin-lattice and spin-spin relaxation rates are found much smaller for the non-magnetic Co$^{3+}$ sites than for the magnetic sites on which the holes are delocalized. The atomic ordering of the Na layers is therefore at the source of this ordered distribution of cobalt charges. The method used here to resolve the Na ordering and the subsequent Co charge order can be used valuably for other concentrations of Na.
\end{abstract}



\maketitle

\section{Introduction}

Since the discovery of high thermo-electric power (TEP)\cite{TEP} and
superconductivity\cite{SC} in Na cobaltates, extended efforts have been done
in order to understand their magnetic and electronic properties. Anomalous
magnetic properties have been discovered in unexpected ranges of Na
concentrations, with an abrupt change of magnetic correlations from AF for $%
x\leq 0.62$ to ferromagnetic for $x\geq 0.67$ \cite{AFandFM}. However, a
situation quite unusual in solid state physics prevailed so far, as very few
experiments have allowed to correlate these magnetic properties with Na
ordering, apart for the featureless band insulator for $x=1$ \cite{Na1CoO2,MHJulien},and the very peculiar $x=1/2$ phase \cite{Zandbergen}$^{-}$\cite{Na0.5CoO2Bobroff}.

We have recently reported that a combination of nuclear magnetic resonance
(NMR) and nuclear quadrupolar resonance (NQR) experiments did concur to let
us determine the Na order in the $x=2/3$ phase \cite{EPL2009}, which is a
nearly ferromagnetic metallic phase which does not order magnetically down
to the lowest temperatures \cite{EPL2008}. For this particular phase the
large set of NMR data obtained in the past allowed us, when complemented by
NQR data, to demonstrate finally that the Na order is rather simple. It
results in a differentiation of Co sites into non-magnetic Co$^{3+}$ sites
in which the t$_{2g}$ subshell of electronic levels is filled, and a
metallic Kagome network of Co sites on which the doped holes are delocalized
\cite{EPL2009}.

We perform here a careful Rietveld analysis of the X-ray data which confirms
the structure proposed from NMR and NQR and definitively establishes the $%
x=2/3$ Na content of this phase contrary to the x=0.71 value estimated by
others from chemical analyzes of single crystal samples \cite{ChouPRL}.
Furthermore, we report in detail here the NQR data which, together with the
approach developed in Ref.~\onlinecite{EPL2009}, will help in similar
analyzes for other structurally more complicated cases detected for
different $x$ values.

The paper is organized as follows. In section II we display the methods used
to synthesize single phase samples for $x=2/3$ and display the powder X-ray
spectra in which the Bragg peaks associated with the Na ordered substructure
are resolved. The best Rietveld fits allowed us to determine the slight
atomic displacements of the Na atoms in the model unit cell deduced from
NMR/NQR data. In section III we report the Na and Co NQR spectra and show
that we could detect most of the NQR lines associated with 4 Co sites
identified by the former combination of NQR and NMR data. Some of the low
frequency lines were however difficult to detect owing to experimental
limitations. At the end of section III we show that the accurate electric
field gradient (EFG) parameters can be hardly described merely by a point
charge model, even taking into account the site displacements determined by
X-rays. This reveals the need for more sophisticated and accurate
calculations taking fully into account the atomic and electronic band
structure. The spin-spin and spin-lattice relaxation data reported in
section IV for the four Co sites are studied and analyzed then, which
allowed us to establish that both quantities can be used to distinguish the
non-magnetic cobalts from those of the more magnetic Co sites which are
organized in a Kagome sublattice in the $x=2/3$ phase.

\begin{figure*}[tbp]
\includegraphics[width=1.0\linewidth]{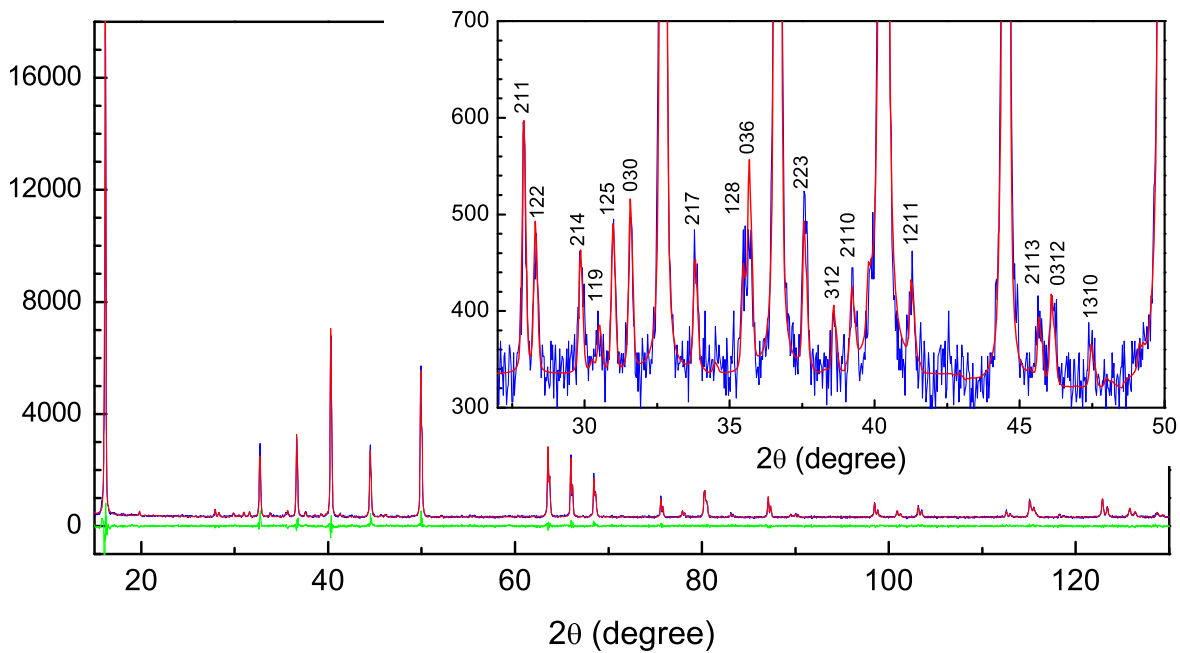}
\caption{(color online) Na$_{2/3}$CoO$_{2}$ structure determination. Main
field: X-ray diffraction pattern observed $I_{obs}$ in blue, calculated $%
I_{calc}$ in red and $I_{obs}-I_{calc}$ in green. The intense 002
substructure reflection at 2$\protect\theta \approx $ 16$^{\circ }$ is not
used for the refinement calculation because of its strong instrumental
asymmetry. Insert: Intensities of satellite reflections: blue observed, red
calculated.}
\label{fig:XRay}
\end{figure*}

\section{Sample preparation and lattice structure}

It has been established that the phase diagram of Na cobaltates is
discontinuous and consists of a series of homogeneous phases extending over
limited concentration ranges, separated by composition gaps \cite%
{AFandFM,EPL2008}. A homogeneous domain range is characterized by a specific
Na ordering which one can monitor by powder X-ray diffraction experiments.
These diffraction patterns were taken at 300~K in the range of $2\theta$
angle $10-130^{\circ }$ (Cu K$\alpha $ radiations were used). X-ray spectra
such as that displayed in Fig.~\ref{fig:XRay} allow to determine the lattice
parameters from the X-ray diffraction peaks of the hexagonal structure of
the CoO$_{2}$ substructure. Also, as shown in the inset of Fig.~\ref%
{fig:XRay}, a set of satellite reflexions associated with the specific Na
ordered structure could be detected and were found quite distinct for the
various stable phases.

\subsection{Sample synthesis}

We have evidenced that in pure oxygen atmosphere the compounds with $x=2/3$
are stable in the temperature range 600-700$^{\circ }$C. Phases with higher
Na content are stabilized at higher temperatures, which for a nominal
composition $x=2/3$ releases some Co$_{3}$O$_{4}$. Conversely lower
preparation temperatures favor phases with $x<2/3$, and in such case the
excess Na is somewhat difficult to detect by diffraction, although after
treatment in air the characteristic diffractions of Na$_{2}$CO$_{3}$ appear.
So to retain the $x=2/3$ phase in powders, the samples were directly
introduced in a furnace stabilized at 700$^{\circ }$C, and were then
quenched at room temperature after treatment. Three different approaches
have been used to synthesize single phase materials:

1) direct synthesis from a stoichiometric composition of Co$_{3}$O$_{4}$ and
Na$_{2}$CO$_{3}$. However we found that the reaction is rather slow at such
low temperatures and requires weeks of treatment with intermediate grindings
to eliminate the last traces of unreacted Co$_{3}$O$_{4}$.

2) from a mixture of cobaltates with calibrated compositions synthesized
previously (such as Na$_{1/2}$CoO$_{2}$ and Na$_{0.71}$CoO$_{2}$\cite%
{EPL2008} taken in a proper ratio.

3) by de-intercalation of Na from Na$_{0.71}$CoO$_{2}$ by annealing it at 700%
${^{\circ}}$C - out of its own stability temperature range. Of course, in
that case the composition $x=2/3$ of the material prepared differs from its
nominal composition $x=0.71$, which remains unchanged and would be detected by chemical analysis. The released Na remains in excess and is not well crystallized (oxidized) and difficult to observe by X-rays.

Overall the second method was found the most reliable and was easily
reproducible.

\subsection{Rietveld refinement of the structure}

\begin{figure}[tbp]
\includegraphics[width=1\linewidth]{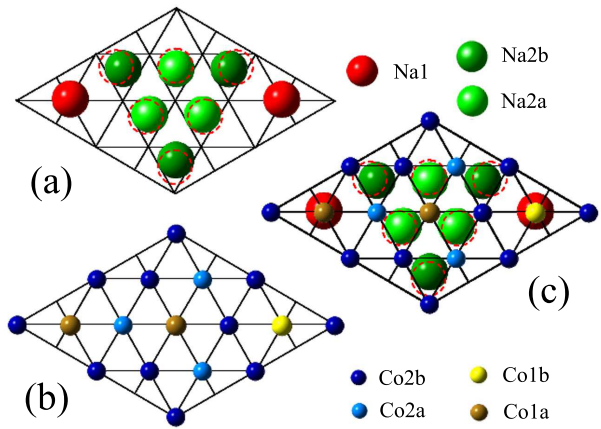}
\caption{Na$_{2/3}$CoO$_{2}$ structure model. Two neighboring sodium and
cobalt layers are shown in one unit cell size. The upper plane (a)
illustrates the Na ordering and the lower one (b) the differentiation in the
Co plane. The dotted circles represent the displaced positions of the sodium
ions, as deduced from the X-ray data analysis. The diameters of the Na and
Co ions scale with the atomic radii of the corresponding ions. (c) Top view
of the two considered Co and Na planes.}
\label{fig:Structure1}
\end{figure}

As already indicated in a previous published report\cite{EPL2009}, the NMR
and NQR data have allowed us to identify 3 Na and 4 Co sites, their
occupancy and those with axial symmetry in the lattice. It has been then
possible to establish the lattice cell of the structure displayed in Fig.~%
\ref{fig:Structure1}. Such a structure ought to be recovered by X-ray
scattering experiments and should explain the set of satellite diffractions
detected, such as those displayed in the inset of Fig.~\ref{fig:XRay}.

For the phases with $x>0.5$ we have detected such sets of satellites, which
in most cases can be described by a single incommensurate wave
vector with component $q_{b}^{\ast}$. Under these conditions the lattice
loses the hexagonal symmetry and becomes generally orthorombic with $%
a_{ort}=a_{hex}\surd 3$, $b_{ort}=a_{hex}/q_{b}^{\ast }$, $c_{ort}=c_{hex}$,
as was noticed early by Zandbergen \textit{et al.} from high resolution
microscopy experiments \cite{Zandbergen}. This has led to propose a
phenomenological model \cite{AFandFM} in which an increase of Na content
corresponded to the insertion of "stripes" of Na2 with increasing $a_{ort}$
in the known structure of Na$_{1/2}$CoO$_{2}$. It is easy to evaluate that
in such a model the empirical relation $x=1-q_{b}^{\ast}$ applies.

In contrast only the particular composition $x=2/3$ exhibits a commensurate
locking leading to a conventional superstructure with a rhombohedral unit
cell $a_{rh}=2a_{hex}\surd 3$, $c_{rh}=3c_{hex}$ which, in the
phenomenological description, corresponds to $q_{b}^{\ast }=1/3$, as
expected. This structure, which corresponds to the spatial group R-3c (n$%
^{\circ }$167) allowed excellent quality Rietveld fits of the X-ray
diffraction pattern as can be seen in the expanded diagrams of Fig.~\ref%
{fig:XRay} \footnote{%
We did not anticipate initially such a large 2D unit cell initially proposed
by Hinuma \emph{et al.} \cite{Hinuma}, and searched a twice smaller
commensurate lattice structure with $a_{ort}=a_{hex}\surd 3$, $%
b_{ort}=3a_{hex}$, $c_{ort}=3c_{hex}$ \cite{EPL2008}. Although in this model
one could index the most significant satellite lines correctly, the fit of
intensities was not as satisfactory as with the complete substructure used.}.
These Rietveld refinements lead to the distribution and occupancy of cobalt
and sodium sites given in Table~\ref{tab:XRay}, in perfect agreement with
the NMR-NQR results. Moreover, they definitively confirm the composition $%
x=2/3$. Indeed, only the sodium sites mentioned in the Table~\ref{tab:XRay}
are \textit{fully occupied} whereas the other sites deduced from the
hexagonal substructure reveal themselves to be completely empty within the
limits of standard deviations.

Most of the ionic coordinates are unchanged with respect to the substructure
positions (see Table~\ref{tab:XRay}). The most displaced are the Na2 ions.
The displacement of Na2 sites already were mentioned in the Ref.~%
\onlinecite{Jorgensen,Cava} but it was attributed to the repulsion of randomly located neighboring Na ions, locally violating the hexagonal symmetry. But as one can see on Fig.~\ref{fig:Structure1} the relaxation of position of these
sites corresponds to a dilatation of the Na2 triangles, which agrees with
the interionic repulsion in this highly charged region. As might be
expected, the displacements of Na along the honeycomb lattice (ionic
conduction channel) of the outer Na2b triangle are nearly twice larger than
those of the inner Na2a triangle.

In addition we would like to mention that we have calculated the diffraction
spectra for eight distinct samples and found in all cases that the Co1a
position is slightly shifted along the c-axis direction towards the Na2
trimers, with the same offset $\approx $ 0.06(1)~\AA. The oxygen ions are
only found slightly displaced with respect to the ideal unit cell, which
points the compactness of the CoO$_{2}$ slabs between which are inserted the
Na ions.

\begin{table}[tbp]
\caption{X ray determination of the structure of Na$_{2/3}$CoO$_{2}$, data
collection 2$\protect\theta $ = 10-130$^{\circ }$, Cu K$_{\protect\alpha }$.
$a=$9.8007(1)\AA ; $3c=$32.8151(4)\AA ; spatial group R-3c (n$^{\circ }$%
~167), hexagonal axes, Z = 72, R = 5.41\%, R$_{w}$ = 7.16\%.}%
\begin{ruledtabular}
\begin{tabular}{lccccc}
Site    & Position\footnotemark[1] & x & y & z & B\\
\hline
Co1b     &  6 b & 0 & 0 & 0 & 0.34(2)\\
Co1a     & 12 c & 0 & 0 & 0.3315(3) & ``\\
Co2a     & 18 d & 1/2 & 0 & 0 & ``\\
Co2b     & 36 f & 1/3\footnotemark[2] & 1/6\footnotemark[2] & 0\footnotemark[2] & ``\\
Na1      & 12 c & 0 & 0 & 1/12\footnotemark[2] & 0.52(7)\\
Na2a     & 18 e & 0.632(2) & 0 & 1/4 & ``\\
Na2b     & 18 e & 0.187(2) & 0 & 1/4 & ``\\
Ox-1     & 36 f & 1/6\footnotemark[2] & 0\footnotemark[2] & 0.0298(5) & 1.3(2)\\
Ox-2     & 36 f & 2/3\footnotemark[2] & 0\footnotemark[2] & 0.0290(4) & ``\\
Ox-3     & 36 f & 1/2\footnotemark[2] & 1/6\footnotemark[2] & 0.0320(5) & ``\\
Ox-4     & 36 f & 1/6\footnotemark[2] & 1/2\footnotemark[2] & 0.0295(5) & ``\\
\end{tabular}
\end{ruledtabular}
\footnotetext[1]{Number of sites and Wyckoff notation for position. All sites are found to be fully occupied (see text).}
\footnotetext[2]{Those coordinates have been left free but
the positions deduced from the fits did not deviate significantly within
experimental accuracy with respect to the mean unit cell positions.}
\label{tab:XRay}
\end{table}

\section{NQR spectra and EFG parameters}

A nucleus with nuclear spin $I>1/2$ has an electric quadrupole moment in
addition to its nuclear magnetic moment. The nucleus interacts with the
electronic environment not only through magnetic hyperfine couplings due to
its magnetic moment, but also through the interaction of its quadrupole
moment with the local crystal electric field gradient (EFG), either static
or dynamic. The EFG arises from a non-symmetric distribution of electric
charge around the nucleus. This charge can originate from non-bonding
electrons, electrons in the bonds and charges of neighboring atoms or ions.
Therefore NQR is a sensitive tool for studying solids as it provides
detailed information on the static and the dynamic properties of the
structure on the scale of a few interatomic spacings. Thus, NQR may be
regarded as a powerful tool for investigating the local order in solids
whereas the interpretation of the data on disordered (or complex-ordered)
materials from usual scattering experiments, such as X-ray or neutron
scattering, is complicated by the absence of the long-range order
translational symmetry or large and complicated unit cell.

Therefore to demonstrate these advantages of NQR to the non-NMR specialist
we start this section by a brief NQR background introduction (A) and
describe in (B) the basic features of the experimental techniques used. Next
we show the $^{23}$Na NQR spectrum (C) and compare it to the $^{23}$Na NMR
data \cite{NaPaper}. We present then (D) the $^{59}$Co NQR spectrum and show
how fast and slow relaxing cobalt sites can be separated in the spectrum.
Finally in (E) we show that the accurate EFG parameters can hardly be
described merely by a point charge model, even taking into account the site
displacements determined by X-rays.

\subsection{NQR Background}

If the crystal field symmetry is lower than cubic, the Hamiltonian of the
quadrupolar interaction can be written\cite{Abragam}:

\begin{equation}
\mathcal{H}_{Q}=\frac{h\nu _{Q}}{6}(3I_{Z}^{2}-I(I-1)+\eta
(I_{X}^{2}-I_{Y}^{2})),  \label{eq:Hamiltonian}
\end{equation}

where the quadrupole frequency
\[
\nu _{Q}=\frac{3eQV_{ZZ}}{2I(2I-1)h}
\]%
is defined by the nuclear quadrupole moment $Q$ and the largest principal
axis component $V_{ZZ}\ $of the EFG tensor  and $\eta =(V_{XX}-V_{YY})/V_{ZZ}
$ is the asymmetry parameter (here the principal axis of EFG tensor are
selected such as $\left\vert V_{ZZ}\right\vert \geq \left\vert
V_{YY}\right\vert \geq \left\vert V_{XX}\right\vert $). In NQR experiments
the magnetic transitions between energy levels with $\Delta m=\pm 1$ are
observed.

\subsection{Experimental techniques}

The NQR measurements were done using a home-built coherent pulsed NMR/NQR
spectrometer. NQR spectra were taken \textquotedblleft point by
point\textquotedblright\ with a $\pi /2-\tau -\pi $ radio frequency (RF)
pulse sequence by varying the spectrometer frequency in equal frequency
steps in the range 1.5-15~MHz at 4.2~K. The detailed NQR spectra were then
constructed using a Fourier mapping algorithm \cite{Clark,Bussandri}.

The change of tuning frequency of the NQR probe has been done with a
variable capacitor. However the Q-factor of the probe (and the sensitivity
of the spectrometer) varies considerably with frequency. This limits
seriously the accuracy of absolute intensity measurements of NQR signals.
Relative intensity measurements could only be done accurately for lines
occurring in a narrow frequency range.

Due to the spin-spin relaxation process (which will be discussed in detail
later) the intensity of the observed spin echo NQR signal decreases with
increasing delay $\tau $ between the RF pulses. It is usually impossible to
reduce this time to zero, as the receiver is overloaded by the RF pulse and
only recovers after a minimum time $\tau _{R}$. Furthermore, for short
values of $\tau $ the detected signal is heavily distorted by residual
oscillations (called "ringing") of the probe components and contains
artifacts. To reduce the probe-ringing time a low Q-factor NQR probe has
been used with Q$\approx $ 10-20. So, the signal can only be detected after
a minimum time $\tau _{D}$ ('dead time') which usually increases markedly
with decreasing frequency. In our experiments the minimum practical $\tau$ values varied from $70~\mu s$ at 1.5~MHz to $40~\mu s$ at 9~MHz.

\subsection{$^{23}$Na NQR spectrum}

The nuclear spin energy levels splitted by the EFG are doubly degenerate
(see Eq.~\ref{eq:Hamiltonian}). So for $^{23}$Na for which $I=3/2$ a single
resonance line which corresponds to the $\pm \frac{1}{2}\leftrightarrow \pm
\frac{3}{2}$ transition can be observed at the frequency:

\begin{equation}
\nu = \nu _{Q}\sqrt{1+\frac{1}{3}\eta ^{2}}.%
\label{eq:quadFreq}
\end{equation}

\begin{figure}[tbp]
\includegraphics [width=0.5\textwidth]{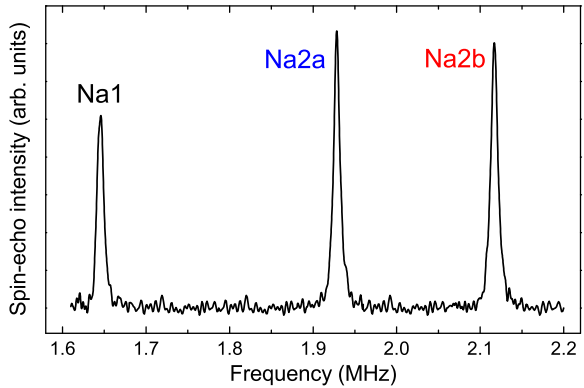}
\caption{$^{23}$Na NQR spectrum in the Na$_{2/3}$CoO$_2$ compound measured
at 4.2~K. The fully resolved quadrupole structure for the three Na sites is
discussed in the text.}
\label{fig:NaSpec}
\end{figure}

The NQR spectrum of $^{23}$Na in the Na$_{2/3}$CoO$_{2}$ compound measured
at 4.2~K is shown in Fig.~\ref{fig:NaSpec}. It consists of three well
resolved narrow lines which correspond to three Na sites with distinct local
environment. Such a spectrum with resolved narrow lines undoubtedly confirms
that this phase is well ordered, as was pointed out in the earlier NMR study
which has established that a characteristic signature of the $x=2/3$ phase
\cite{NaPaper} is to exhibit three distinct sodium sites. The shape of the $%
^{23}$Na NQR lines has been fitted by a Lorentzian function and their
positions and linewidths (which were taken as a full width at half maximum -
FWHM) are collected in  Table~\ref{tab:NaParams}.

As for $I=3/2$ the NQR frequency determined by Eq.~(\ref{eq:quadFreq})
depends on both $\nu _{Q}$ and $\eta $, it is impossible to determine both
parameters from the NQR spectrum. However, these parameters were
determined by NMR for each sodium site and are also listed in Table~\ref%
{tab:NaParams}. Using these values and following Eq.~(\ref{eq:quadFreq}) one
can calculate the expected frequencies $f_{calc}$ of $^{23}$Na NQR lines
(Table~\ref{tab:NaParams}). As one can see the calculated and experimental
values are in perfect agreement - therefore the notation proposed in Ref.~%
\onlinecite{NaPaper} for different sodium sites have been used in Table~\ref%
{tab:NaParams}. The tiny ($<$1.3\%) difference in the experimental and
calculated frequencies for the Na2b site can be easily explained by a small
deviation (not more than 5$^{\circ }$) of the Z
principal axis of the EFG tensor and the c crystallographic axis for this
site. The relative intensities of sodium lines will be discussed later in
section IV of this paper.

\subsection{$^{59}$Co NQR spectrum}

Former $^{59}$Co NMR data \cite{CoPaper} taken on this phase have already
allowed us to evidence distinct Co NMR lines, but the NMR spectra were
somewhat difficult to analyze fully, as one needs to determine altogether
the EFG parameters and NMR shifts of the various sites. The experiments were
furthermore complicated by the need of a quasi perfect alignment of the
powder sample with respect to the applied magnetic field. They however
allowed us to determine some of the EFG parameters, which allowed us to know
beforehand the range of expected $\nu _{Q}$ values. After a presentation of
the spectra, we explain in some detail how the various transitions
pertaining to the same sites can be identified. The identification of the Co
NQR lines is explained in some details, and allow us then to determine the
EFG values.

\begin{table}[tbp]
\caption{Comparison of $^{23}$Na NQR and NMR results. $\nu_{calc}$ is the
calculated value for the NQR resonance line frequency using Eg.~(\ref{eq:quadFreq}) for the values of $\nu_{Q}$ an $\eta$
obtained by NMR (see text).}
\label{tab:NaParams}%
\begin{ruledtabular}
\begin{tabular}{lccccc}
&\multicolumn{2}{c}{NQR}&\multicolumn{2}{c}{NMR\cite{NaPaper}} &\\
Site & $\nu$(MHz) & $\Delta \nu$ (kHz) & $\nu_Q$(MHz) & $\eta$ & $\nu_{calc}$(MHz)\\
\hline
Na1  & 1.645(1) & 8.2(1) & 1.645(5) & 0.01(1) & 1.645 \\
Na2a & 1.928(2) & 8.2(1) & 1.74(1)  & 0.84(2) & 1.93  \\
Na2b & 2.117(1) & 9.0(1) & 1.86(1)  & 0.89(2) & 2.09  \\
\end{tabular}
\end{ruledtabular}
\end{table}

\subsubsection{Experimental results}

\begin{figure*}[tbp]
\includegraphics[width=0.8\textwidth]{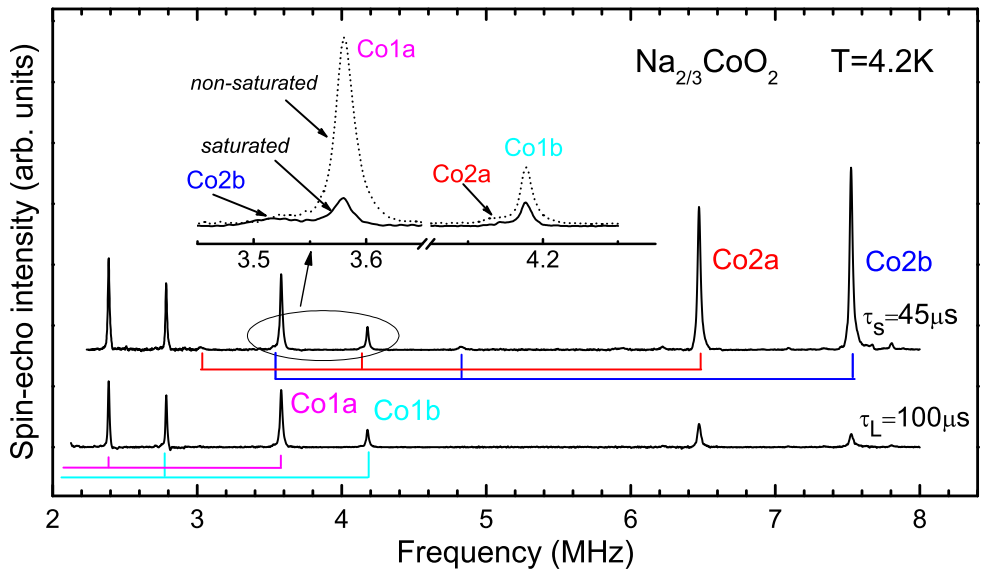}
\caption{Main panel: $^{59}$Co NQR spectra taken at 4.2~K for short ($%
\protect\tau _{S}=45~\protect\mu s$) and long ($\protect\tau _{L}=100~%
\protect\mu s$) delay between pulses. Insert: the two weak intensity lines
of the $\pm $3/2$\longleftrightarrow \pm $5/2 Co2a and Co2b sites are
revealed by saturation of the high intensity slow relaxing lines.}
\label{fig:CoSpec}
\end{figure*}

Typical NQR spectra of $^{59}$Co in the Na$_{2/3}$CoO$_{2}$ compound are
shown in Fig.~\ref{fig:CoSpec}. We found that the number of lines and their
intensities in the observed spectrum depend strongly on the delay $\tau$
between pulses. This corresponds to the fact that the experimental spin-echo
intensity depends on the rate of the spin-spin relaxation process.
Consequently the fast relaxing nuclei are not observable at long enough
delay $\tau$ between pulses, as will be detailed in section IV. In the
spectrum observed with the shortest possible $\tau_{S}=45~\mu s$ there are 6
narrow intense lines and several lines with weaker intensity (Fig.~\ref%
{fig:CoSpec}). At the same time in the spectrum measured with $\tau
_{L}=100~\mu s$ the intensities of the high frequency lines greatly
decrease, while the intensities of 4 low frequency narrow lines does not
change a lot. So, it is obvious that in this phase with $x=2/3$ two
different types of Co sites - fast and slow relaxing - coexist, as was
established already by NMR\cite{CoPaper}.

As was shown earlier \cite{EPL2008,AFandFM} for the high sodium content
range ($x>0.5$) of the sodium cobaltates phase diagram, a cobalt charge
disproportionation into Co$^{3+}$ and Co$^{\approx 3.5+}$ is a quite common
picture. Cobalt ions in sodium cobaltates are in low spin configurations, so
Co$^{3+}$ has an electronic spin $S=0$ and relax rather slowly in comparison
with the cobalts with higher charge state on which holes delocalize
(formally Co$^{4+}$ should have $S=1/2$).

\subsubsection{Identification of the NQR lines}

\begin{figure}[tbp]
\includegraphics[width=1.0\linewidth]{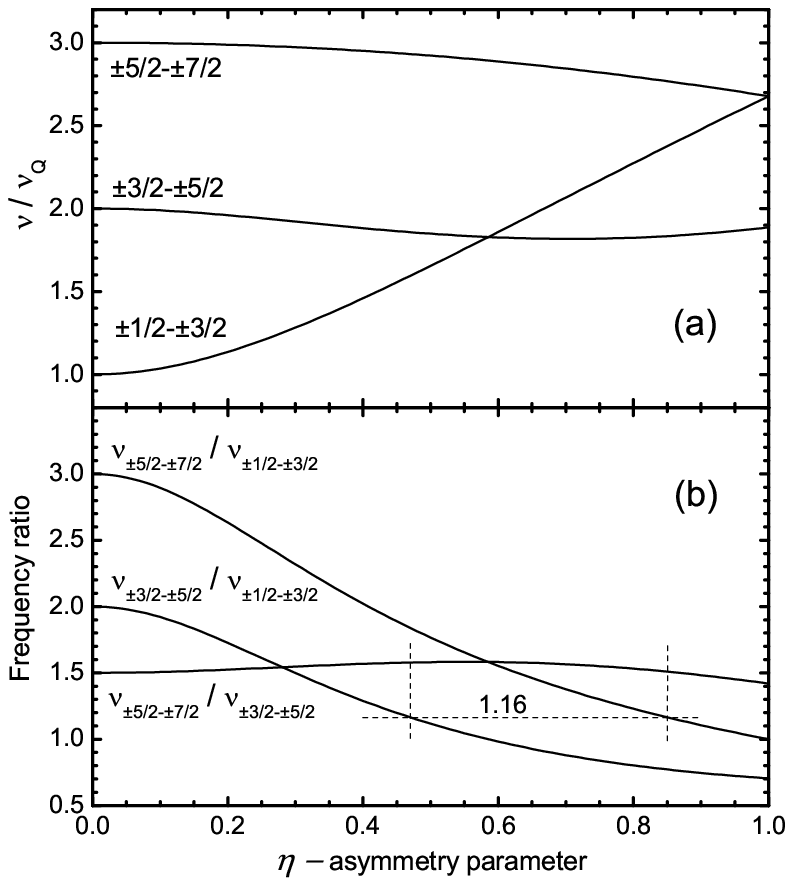}
\caption{(a) Theoretical dependence of the allowed NQR transition frequencies on the asymmetry parameter $\eta$ of the EFG tensor for a nuclear spin $I=7/2$. (b) Frequency ratios of the allowed transitions plotted versus $\eta$. The particular case of the ratio 1.16 used in the text to illustrate the analysis is shown by dotted lines.}
\label{fig:NQRTheory}
\end{figure}

As the $^{59}$Co nuclei have $I=7/2$, for a single crystallographic site one
should observe up to 3 lines in the NQR spectrum. The positions of these
lines depends on both $\nu _{Q}$ and $\eta $. Figure~\ref{fig:NQRTheory}a
shows the theoretical dependence of NQR frequencies of the allowed
transitions on the asymmetry parameter of the EFG tensor for nuclear spin
7/2. These data were obtained by numerical diagonalization of the main
Hamiltonian recalled in Eq.~(\ref{eq:Hamiltonian}). One can see that the
frequencies of the $\pm \frac{3}{2}\leftrightarrow \pm \frac{5}{2}$ and $\pm
\frac{5}{2}\leftrightarrow \pm \frac{7}{2}$ transitions weakly depend on the
asymmetry parameter $\eta $, contrary to that of the $\pm \frac{1}{2}%
\leftrightarrow \pm \frac{3}{2}$ transition (Fig.~\ref{fig:NQRTheory}a). For
$\eta =1$ only two resonance lines in the spectrum should be observed.
Measuring experimentally the resonance frequencies for different transitions
should allow then to determine both $\nu _{Q}$ and $\eta $ for $I=7/2$.

Since the number of lines in the experimental $^{59}$Co NQR spectrum (Fig.~%
\ref{fig:CoSpec}) is much larger than 3 (number of the observable NQR lines
for $I=7/2$), an analysis has to be done to determine the triplet of lines
which are associated with a given cobalt site in the unit cell of this $x=2/3
$ compound. The basic data which can be used for such an analysis are the
theoretical ratios of frequencies of different transitions for a single
cobalt site which are restricted as established from Fig.~\ref{fig:NQRTheory}%
b.

Let us consider as an example the two intense lines at $\approx $7.52~MHz
and $\approx $6.47~MHz (Fig.~\ref{fig:CoSpec}) which corresponds to an an
experimental frequency ratio $\approx $1.16. Although we know from NMR
experiments \cite{CoPaper} that those EFG values correspond to distinct Co
sites, we want to show here that a simple analysis of the NQR spectra does
confirm that point. In theory such a ratio could be found for a single site
with EFG's with $\eta =0.47$ or $\eta =0.85$ (see Fig.~\ref{fig:CoSpec}b).
As $\eta =0.47$ corresponds to the ratio of frequencies of $\pm \frac{3}{2}%
\leftrightarrow \pm \frac{5}{2}$ and $\pm \frac{1}{2}\leftrightarrow \pm
\frac{3}{2}$ transitions, then one should find in the spectrum a line
corresponding to the $\pm \frac{5}{2}\leftrightarrow \pm \frac{7}{2}$
transition at a frequency $\approx 7.52\ast 1.578\approx 11.87$~MHz. We
indeed didn't find such a line in our experiment. Also we know from NMR that
such large EFG values do not exist \cite{CoPaper}. Similarly for $\eta =0.85$
one should find a line for the $\pm \frac{3}{2}\leftrightarrow \pm \frac{5}{2%
}$ transition at $7.52/1.51\approx 4.98$~MHz which does not exist either.
Therefore the $\approx $7.52~MHz and $\approx $6.47~MHz lines cannot be
attributed to the same cobalt site and correspond to two different fast
relaxing cobalt sites Co2b and Co2a.

Then one has to consider the next pair of lines, e.g. at $\approx $7.52~MHz
and $\approx $4.83~MHz. The frequency ratio is equal $\approx $1.56 and 3
possible values of $\eta $ should be considered. However only one value of $%
\eta \approx $ 0.36 gives the right 3rd line position in the spectrum, a
weak and broad line at $\approx $3.52~MHz which completes the spectrum of
the Co2a site. We found this line by observing the slight asymmetry of the shape of the intense and narrow slow relaxing line at 3.581~MHz decreases with increasing $\tau$. To better resolve this line we used the fact that slow and fast-relaxing cobalts can be separated also by the large difference in their spin-lattice relaxation rates, as seen by NMR and as detailed later in section IV. To saturate the signal an additional $\pi /2$ pulse with 100$\mu s$ delay was used before the $\pi /2-\tau -\pi $ pulse sequence. In the spectrum obtained after such a pulse sequence ("saturated" in the insert of Fig.~\ref{fig:CoSpec}), the intensities of the slow-relaxing cobalt lines decrease and the lines of the fast-relaxing cobalts can be better resolved. This method allowed us to detect one more fast relaxing line at 4.145~MHz which completes the three line spectrum for the 2nd fast relaxing Co2a site, with the two other lines at 6.473~MHz and 3.03~MHz.

Such an analysis allowed us also to determine that the two slow-relaxing
lines at 3.581 and 4.178~MHz are the $\pm \frac{5}{2}\leftrightarrow \pm
\frac{7}{2}$ transitions of two different sites of cobalt, Co1a and Co1b, as
could already be anticipated from the Co NMR data \cite{CoPaper}. The
corresponding $\pm \frac{3}{2}\leftrightarrow \pm \frac{5}{2}$ transitions
are at 2.387 and 2.785~MHz. We did not attempt to observe the $\pm \frac{1}{2%
}\leftrightarrow \pm \frac{3}{2}$ transitions for these slow-relaxing
cobalts which should appear at low frequencies $\approx $~1.19 and $\approx $%
~1.39~MHz, outside of the frequency range of our spectrometer.

\subsubsection{$^{59}$Co EFG parameters}

All the observed $^{59}$Co NQR lines were fitted by Lorentzian functions and
their positions and linewidths are collected in Table~\ref{tab:CoNQR} and
Table~\ref{tab:CoRelax}, respectively. The values of $\nu _{Q}$ and $\eta $
found in this work are in perfect agreement with those quoted in the Ref.~%
\onlinecite{CoPaper}. The fact that we observe a limited number of narrow
and well-resolved $^{59}$Co NQR lines (Fig.~\ref{fig:CoSpec}) confirms the
existence of long-range order in this $x=2/3$ phase of sodium cobaltates.
Moreover NQR is clearly able to distinguish the charge segregation between
the slow relaxing non-magnetic Co$^{3+}$ (sites Co1a and Co1b), and the fast
relaxing cobalt sites Co2a and Co2b, on which holes are delocalized. Despite
all our efforts we have not found in the $^{59}$Co NQR spectrum any traces
of the Co3 site anticipated from NMR (see Ref.~\onlinecite{CoPaper}). As
explained in the Ref.~\onlinecite{EPL2009} we did conclude that this
apparent "site" was an experimental artifact due to imperfect alignment of
the NMR sample. This point will be detailed in an extensive experimental
report of the Co NMR spectra \cite{H67_CoNMR}. The intensities of the $
^{59}$Co NQR lines will be discussed later in section IV after the analysis
of the spin-spin relaxation of these cobalt NQR signals.

\begin{table}[tbp]
\caption{Measured frequencies for the NQR transitions of the four Co sites in Na$_{2/3}$CoO$_{2}$ and deduced values of $\nu _{Q}$ and $\eta$.}
\label{tab:CoNQR}%
\begin{ruledtabular}
\begin{tabular}{ccccccc}
& \multicolumn{3}{c}{Frequency(MHz)}& & &\\
Site & $\pm\frac{1}{2}\leftrightarrow \pm\frac{3}{2}$ & $\pm\frac{3}{2}\leftrightarrow \pm\frac{5}{2}$ &
$\pm\frac{5}{2}\leftrightarrow \pm\frac{7}{2}$
& & $\nu_Q$(MHz) & $\eta$ \\
\hline
Co1a & -       & 2.387(1) & 3.581(2) & & 1.193(1) & $\leq$0.017 \\
Co1b & -       & 2.785(1) & 4.178(1) & & 1.392(1) & $\leq$0.016 \\
Co2a & 3.03(1) & 4.145(3) & 6.473(1) & & 2.187(1) & 0.362(5) \\
Co2b & 3.52(1) & 4.826(2) & 7.524(1) & & 2.541(1) & 0.358(4) \\
\end{tabular}
\end{ruledtabular}
\end{table}

\subsubsection{NQR linewidths}

\begin{figure}[h]
\includegraphics[width=1.0\linewidth]{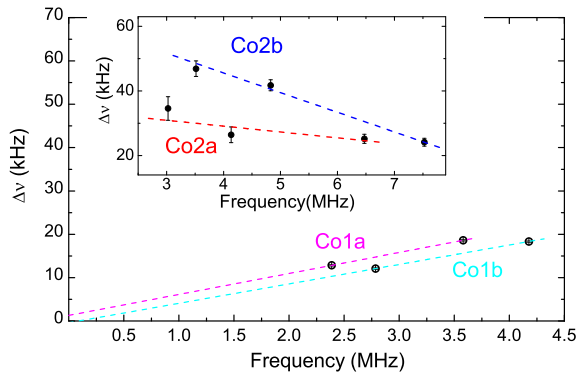}
\caption{Lines widths at half height for the NQR transitions of the
different sites. The main panel is for Co1a,b and inset for Co2a,b sites.}
\label{fig:lineWidth}
\end{figure}

The broadening of the Co1 and Co2 groups of lines were found to display
different behaviours likely linked with their different magnetic properties.
Figure~\ref{fig:lineWidth} shows the linewidths variation with resonance
frequency of the various transitions for each of the cobalt NQR lines. Two
different behaviors appear clearly: the linewidths increase with increasing
frequency for Co1 sites while the opposite holds for the Co2 sites. The Co1a
and Co1b sites are non-magnetic with localized holes and one expects
essentially a broadening due to a distribution of EFG values. Indeed for a
spin 7/2 system with zero asymmetry parameter (Co1 case), a spread $e\Delta q
$ of EFG values induces broadenings of the different NQR transitions such
that $\Delta \nu _{i}/\nu _{i}\propto \Delta q/q$, so that the linewidth
increases linearly with the frequency of the transition as found in the data
of Figure ~\ref{fig:lineWidth}. One would expect then a linewidth of $%
\approx $ 6~kHz for the lowest ($\pm \frac{1}{2}\leftrightarrow \pm \frac{3}{%
2}$) transitions of the Co1a,b sites. Extrapolating the linear fits for
Co1a,b sites to zero frequency gives a very small residual value $\approx $
1~kHz, which could be attributed to dipole dipole coupling as will be
detailed in section IV.

As Co2a and Co2b are magnetic sites, a broadening due to magnetic
interactions might take place and apparently increases with decreasing index
of the quadrupole transition. This will be discussed in section IV after
considering the spin spin $T_{2}$ processes.

\subsection{Rough estimates of the EFG values}

We have calculated the EFG tensors and the corresponding parameters of the
quadrupole Hamiltonian ($\nu _{Q}$ and $\eta $) for all sites of the $^{23}$%
Na and $^{59}$Co in the simplest point charge model assuming charges 1$^{+}$
on Na, 3$^{+}$ on Co1, 3.44$^{+}$ on Co2 and 2$^{-}$ on O  ( see Table~\ref%
{tab:EFGcalc}). Of course one does not expect such point charge calculations
to fit perfectly the data as similar calculations do not even explain the
simple case of Na$_{1}$CoO$_{2}$ for which all Co are Co$^{3+}$ (Ref.~%
\onlinecite{Na1CoO2} and references therein). Such an approach does not take
into account correctly the cobalt-oxygen covalency. We do however clearly
see in the results displayed in the Table~\ref{tab:EFGcalc} that for the Na1
and Co1 sites the asymmetry parameter is of course that expected from the
site symmetry, that is $\eta =0$. We also find that the asymmetry of the
structure yields rather large asymmetries for the EFG values on the Na2 and
Co2 sites as found experimentally. However the magnitudes of the EFG
computed do not agree quantitatively with the experimental values. We did
however find that the displacement of the Na atoms found in the X-ray
structure analysis yields significant changes in the calculated EFG values.
This reveals that more sophisticated calculations taking into account the
atomic displacements and the band structure at least in the local density
approximation (LDA) are required to explain the EFG parameters.

\begin{table}[tbp]
\caption{Results of the point charge model calculations of the EFG
parameters for established structure of the Na$_{2/3}$CoO$_2$. All
quadrupolar frequencies $\nu _{Q}$ are in MHz. $\theta$ (in degree) is the angle between the Z principal axis of the EFG tensor and the \emph{c} crystallographic axis. Na: $^{23}$Q=0.1; $^{23}\gamma $=-4.1; Co: $^{59}$Q=0.42; $^{59}\gamma $=-7;}
\label{tab:EFGcalc}%
\begin{ruledtabular}
\begin{tabular}{cccc}
Site&Experiment&Non-shifted&shifted\\
  \hline
  \hline
$ $ & $\nu_Q$=1.645(5) & $\nu_Q$ = 0.61 & $\nu_Q$ = 0.28 \\
Na1 & $\eta =0$ & $\eta=0$ & $\eta = 0$\\
$ $ & $\theta =0$ & $\theta =0$ & $\theta=0$\\
\hline
$ $ & $\nu_Q$=1.74(1) & $\nu_Q$ = 0.93 & $\nu_Q$ = 0.94\\
Na2a & $\eta =0.84(2)$ & $\eta =0.88$ & $\eta = 0.19$\\
$ $ & $\theta =0$ & $\theta =90$ & $\theta=90$\\
\hline
$ $ & $\nu _{Q}$=1.86(1) & $\nu_Q$ = 0.53 & $\nu_Q$ = 1.15\\
Na2b & $\eta =0.89(2)$ & $\eta=0.31$ & $\eta = 0.36$\\
$ $ & $\theta \leq 5$ & $\theta=90$ & $\theta=90$\\
\hline
$ $ & $\nu _{Q}$=1.193(1) & $\nu _{Q}$ = 1.06& $\nu _{Q}$ = 1.71\\
Co1a & $\eta \leq 0.017$ & $\eta=0$ & $\eta = 0$\\
$ $ & & $\theta =0$ & $\theta=0$\\
\hline
$ $ & $\nu _{Q}$=1.392(1) &$\nu _{Q}$ = 1.18&$\nu _{Q}$ = 0.94\\
Co1b & $\eta \leq 0.01b$ & $\eta=0$ & $\eta = 0$\\
$ $ & & $\theta =0$ & $\theta=0$\\
\hline
$ $ & $\nu _{Q}$=2.187(1) & $\nu_Q$ = 1.42&$\nu_Q$ = 1.55\\
Co2a & $\eta =0.362(5)$ & $\eta=0.26$ & $\eta=0.56$\\
$ $ & $ $ & $\theta=18$ & $\theta=14$\\
\hline
$ $ & $\nu_Q$=2.541(1) & $\nu_Q$ = 1.55 & $\nu_Q$ = 1.37\\
Co2b & $\eta =0.358(4)$ & $\eta=0.56$ & $\eta=0.40$\\
$ $ & $ $ & $\theta=11$ & $\theta=12$\\
\end{tabular}
\end{ruledtabular}
\end{table}

\section{Spin-lattice relaxation and Spin-spin relaxation}

When thermal equilibrium of the nuclear spins is disturbed by RF pulses,
their equilibrium magnetization is recovered by the nuclear nuclear
spin-lattice relaxation (NSLR) process which corresponds to the relaxation
of the longitudinal component of the magnetization. The decay of the
transverse component of the magnetization,  the nuclear spin-spin relaxation
(NSSR) process  is connected with the loss of  coherence inside the spin
system due to spin-spin interactions. Generally both relaxation processes
could have different origins which reflect magnetic and electronic
properties of the materials, the inner arrangement, the interactions in the
spin system, the different motion and diffusion processes. In this section
we report and discuss first in (A) the $^{59}$Co spin-lattice relaxation for
all four cobalt sites. Next we consider in (B) the spin-spin relaxation starting from $^{23}$Na as a simple case of NSSR. Then we analyze the $^{59}$Co spin-spin relaxation. We demonstrate that both NSLR and NSSR allow to differentiate the two types of Co sites - non-magnetic Co$^{3+}$ ions on 25\% of the cobalt sites and a metallic Kagome network of Co sites on which the doped holes are delocalized \cite{EPL2009}.

\subsection{$^{59}$Co spin - lattice relaxation}

To study the NSLR process we have used the magnetization inversion
recovery method which uses three pulses: $\pi -t-\pi /2-\tau -\pi $, where
$t$ and $\tau $ are the time intervals between pulses. In this sequence the
first pulse rotates the magnetization by 180 degree, the second and third
pulses gives a spin-echo, and the dependence of the spin-echo intensity on
delay time $t$ allows to monitor the recovery of the nuclear magnetization
associated with a given NQR transition:

\begin{equation}
M(t)=M_{0}(1-B\cdot R(t)). %
\label{eq:Kinetics}
\end{equation}

Here $M_{0}$ is the thermal equilibrium value of magnetization and the
parameter $B$ characterizes the actual magnetization after the first pulse
at $t=0$ (the imperfection of the experimental conditions gave typical
values $B$ $\simeq $ 1.8 rather than $B=2$ expected for a perfect $\pi $
pulse). The shape of the relaxation function $R(t)$ depends on the nuclear
transition sampled. For a two-level nuclear system (like the $I=3/2$ NQR
case) this process is exponential $R(t)=exp\left( -t/T_{1}\right) $, with a
characteristic time constant $T_{1}$, the spin-lattice relaxation time\cite%
{Slichter}. But generally for $I>1/2$ \ the nuclear energy levels are
differentiated by the quadrupole interaction with the crystalline electric
field (Eq.~\ref{eq:Hamiltonian}). As a consequence the difference in
population between adjacent levels which are probed by the RF pulses depends
on the populations of the levels which are not hit by the RF pulses.
Therefore the magnetization recovery becomes multi-exponential
\begin{equation}
R(t)=\sum_{i}a_{i}exp\left( -\frac{\lambda _{i}t}{T_{1}}\right) ,
\label{eq:Recovery}
\end{equation}%
but is still characterized by a single $T_{1}$value.

The spin-lattice relaxation could be driven either by magnetic or
quadrupolar fluctuations. However in the sodium cobaltates, which
exhibits unusual magnetic properties, the magnetic relaxation mechanism
dominates at least at low $T$. Following Ref.~\onlinecite{Chepin} we
obtained the parameters for the theoretical relaxation functions $R(t)$ for
the case of magnetic relaxation by weak isotropic fluctuating magnetic
fields for different transitions of different Co sites - see Table~\ref%
{tab:CoT1Coef}.

\begin{table}[tbp]
\caption{Theoretical spin-lattice relaxation function (\ref{eq:Recovery}) parameters for different $^{59}$Co sites for the case of magnetic relaxation by weak isotropic fluctuating magnetic fields (based on Ref.~\onlinecite{Chepin}).}
\label{tab:CoT1Coef}%
\begin{ruledtabular}
\begin{tabular}{ccccccccc}
Site & $\eta$ & Transition & $a_1$ & $\lambda_1$ & $a_2$ & $\lambda_2$ & $a_3$ & $\lambda_3$ \\
\hline
Co1 & 0 & $\pm \frac{5}{2}\leftrightarrow \pm \frac{7}{2}$ & 0.14 & 21 & 0.65 & 10 & 0.21 & 3 \\
`` & `` & $\pm \frac{3}{2}\leftrightarrow \pm \frac{5}{2}$ & 0.74 & 21 & 0.16 & 10 & 0.1 & 3 \\
Co2 & 0.36 & $\pm \frac{5}{2}\leftrightarrow \pm \frac{7}{2}$ & 0.27 & 17.7 & 0.53 & 8.96 & 0.2 & 3 \\
`` & `` & $\pm \frac{3}{2}\leftrightarrow \pm \frac{5}{2}$ & 0.88 & 17.7 & 0.038 & 8.96 & 0.082 & 3 \\
\end{tabular}
\end{ruledtabular}
\end{table}

\begin{figure}[tbp]
\includegraphics[width=1.0\linewidth]{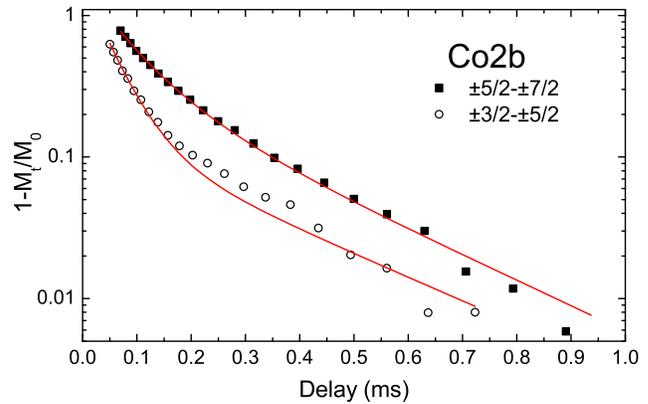} .
\caption{Spin-lattice relaxation curves for Co2b site measured at 7.52~MHz
and 4.83~MHz ($\pm\frac{5}{2} \leftrightarrow \pm\frac{7}{2}$ and $\pm\frac{3%
}{2} \leftrightarrow \pm\frac{5}{2}$ transitions correspondingly) at 4.2~K.
Solid lines are the fits of experimental points by the magnetization
relaxation functions (\ref{eq:Recovery}) with parameters from the
Table~\ref{tab:CoT1Coef}.}
\label{fig:CoT1}
\end{figure}

Figure~\ref{fig:CoT1} shows experimental spin-lattice relaxation curves for
the $\pm \frac{5}{2}\leftrightarrow \pm \frac{7}{2}$ and $\pm \frac{3}{2}%
\leftrightarrow \pm \frac{5}{2}$ transitions for the Co2b site measured at
7.52 and 4.83~MHz, respectively. The fits with Eq. (4) with the $a_{i}\ $and
$\lambda _{i}\ $taken from Table~\ref{tab:CoT1Coef} are shown by solid
lines. These fits give $T_{1}$ values of 0.73(4)~ms and 0.8(1)~ms
respectively for the $\pm \frac{5}{2}\leftrightarrow \pm \frac{7}{2}$ and $%
\pm \frac{3}{2}\leftrightarrow \pm \frac{5}{2}$ (Table~\ref{tab:CoRelax}).
This rather good agreement confirms that the relaxation is caused by
magnetic fluctuations and that the spectral density of fluctuating fields is
nearly independent on the frequency.

For the Co1a and Co1b sites it was found that $T_{1}$ values for the $\pm
\frac{5}{2}\leftrightarrow \pm \frac{7}{2}$ is shorter than that for $\pm
\frac{3}{2}\leftrightarrow \pm \frac{5}{2}$ (Table~\ref{tab:CoRelax}). This
effect does result in that case from the proximity of the fast relaxing Co2 $%
\pm \frac{3}{2}\leftrightarrow \pm \frac{5}{2}$ transitions (see Fig.~\ref%
{fig:CoSpec}) which induce by cross-relaxation processes a reduction of the
the recovery time on the $\pm \frac{5}{2}\leftrightarrow \pm \frac{7}{2}$
transitions of Co1a and Co1b. For these Co1\ sites the $T_{1}$ value
retained then hereafter is the longer one deduced from the $\pm \frac{3}{2}%
\leftrightarrow \pm \frac{5}{2}$ transition data.

\begin{figure}[bp]
\includegraphics[width=1.0\linewidth]{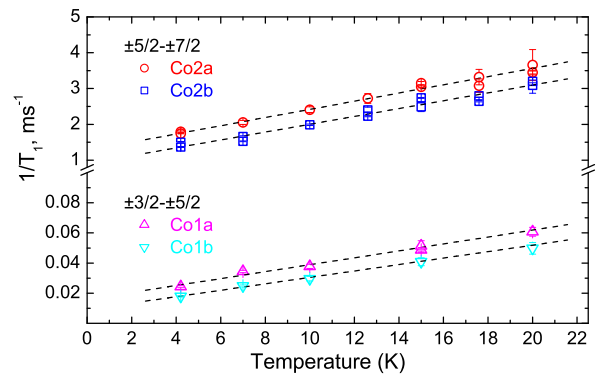}
\caption{Temperature dependence of nuclear spin-lattice relaxation rates $%
1/T_1$ for all four cobalt sites in the Na$_{2/3}$CoO$_2$ compound. For Co1a
and Co1b measurements were done on the $\pm\frac{3}{2}\leftrightarrow \pm
\frac{5}{2}$ transitions and for Co2a and Co2b measurements were done on the
$\pm \frac{5}{2}\leftrightarrow \pm \frac{7}{2}$ transitions. Linear fits shown by dotted lines are guides to the eyes.}
\label{fig:CoT1T}
\end{figure}

In Fig.~\ref{fig:CoT1T} we show the temperature dependence of nuclear
spin-lattice relaxation rates $1/T_{1}$ for all four cobalt sites. As one
can see for both Co2a and Co2b NSLR is about two orders of magnitude larger
than for Co1a and Co1b. As was already stated in Ref.~\onlinecite{EPL2009},
this proves the non-magnetic nature of Co1 sites. Also Fig.~\ref{fig:CoT1T}
shows that $1/T_{1}$ increases with $T$ at low temperatures. Above 20~K the
NSLR becomes so fast that we loose then the Co2 NQR signals. Measurements at higher $T$ cannot be carried out then by NQR, and $T_{1}$ evolution will be rather studied by NMR as will be discussed in a forthcoming paper \cite{H67_CoNMR}. Similarly the spin-lattice relaxation measurements for $^{23}$Na were technically challenging and have therefore not been performed here as they have been studied in great detail by NMR in Ref.~\onlinecite{EPL2008} for various Na$_{x}$CoO$_{2}$ phases including the
$x=2/3$ phase.

\begin{table*}[tbp]
\caption{$^{59}$Co relaxation and NQR linewidth parameters (see text for details).}
\label{tab:CoRelax}%
\begin{ruledtabular}
\begin{tabular}{cccccccc}
Site & Co1a & Co1b & Co1a & Co1b & Co2b & Co2a & Co2b\\
$Transition $ & $\pm\frac{3}{2}\leftrightarrow \pm\frac{5}{2}$ & $\pm\frac{3}{2}\leftrightarrow \pm\frac{5}{2}$ & $\pm\frac{5}{2}\leftrightarrow \pm\frac{7}{2}$ & $\pm\frac{5}{2}\leftrightarrow \pm\frac{7}{2}$ & $\pm\frac{3}{2}\leftrightarrow \pm\frac{5}{2}$ & $\pm\frac{5}{2}\leftrightarrow \pm\frac{7}{2}$ & $\pm\frac{5}{2}\leftrightarrow \pm\frac{7}{2}$ \\
\hline
$\nu$~(MHz)      & 2.387(1) & 2.785(1) & 3.581(2) & 4.178(1) & 4.826(2) & 6.473(1) & 7.524(1)\\
$T_{1}$~(ms) & 40.5(5) & 55.5(5)  & 26(1)    & 46(2)    & 0.8(1)   & 0.59(3)  & 0.73(4) \\
$T_{2}$~($\mu $s)     & 316(4)   & 326(7)   & 320(2)   & 335(6)   & 30(2)    & 71(2)    &  49(2)  \\
$n$          & 2 & 2 & 2 & 2 & 1 & 1 & 1 \\
$T_{1}'$~(ms)     & 2.3 & 3.2 & 2.6 & 4.6 & 0.050 & 0.058 & 0.072 \\
$\alpha$    & - & - & - & - & 0.61(5) & 1.20(8) & 0.67(5) \\
$\Delta \nu _{calc}$~(kHz)    & 1.0(4)  & 0.9(2)  & 1.0(2)  & 0.9(2)  & 10.6(3)   & 4.5(3)  & 6.5(2) \\
$\Delta \nu$~(kHz)   & 12.8(9)  & 12.1(8)  & 18.6(9)  & 18.3(8)  & 42(2)    & 24(1)    & 24(1)   \\
\end{tabular}
\end{ruledtabular}
\end{table*}

\subsection{Spin-spin relaxation and signal intensities}

NSSR is a complex phenomenon, but at its most fundamental level the random
fluctuations of the local magnetic field leads to a loss of the initial
phase coherence of the nuclear spins and therefore to the decrease of the
transverse nuclear spin magnetization. In the Redfield theory\cite{Slichter}
the transverse relaxation arises from two mechanisms: (1) nuclear spin-spin
coupling (via magnetic dipolar or indirect interactions) and (2) the
spin-lattice relaxation.

The NSSR is studied by monitoring the spin-echo intensity as a function of
the time delay $\tau $ between $\pi $/2 and $\pi $ pulses. In general the
decay of transverse magnetization $M(t)$ as a function of time $t=2\tau $
can be fitted by the equation:
\begin{equation}
M(t)=M(0)exp\left( -\left( \frac{t}{T_{2}}\right) ^{n}\right) .
\label{eq:funcT2}
\end{equation}%
The NSSR process is characterized by the relaxation time $T_{2}$ and the
exponent $n$, which characterizes the decay shape which usually varies
between Lorentzian $(n=1)$ and Gaussian $(n=2)$.

\subsubsection{$^{23}$Na spin-spin relaxation}

\begin{figure}[bp]
\includegraphics[width=1.0\linewidth]{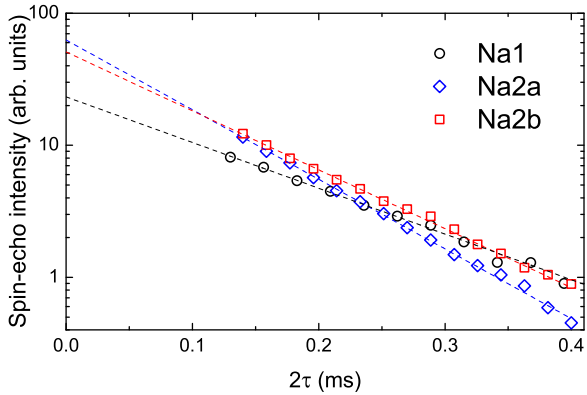}
\caption{Spin-spin relaxation curves of $^{23}$Na sites and fits by the Eq.~(%
\protect\ref{eq:funcT2}).}
\label{fig:NaT2}
\end{figure}

The spin-spin relaxation decays for sodium sites are shown in Fig.~\ref%
{fig:NaT2}. The shortest $\tau $ value used was only $\approx 65~\mu s$ due
to the long 'dead time' $\tau _{D}$ of the spectrometer (see section III).
Then the beginning of the relaxation decays were lost and only tails were
measured. Therefore it was impossible to determine reliably the transverse
magnetization decay shape. Nevertheless the experimental data were fitted by
a Lorentzian $(n=1)$ function (see Eq.~(\ref{eq:funcT2})) in order to obtain
estimates of the spin-spin relaxation times and of the NQR lines
intensities. The $T_{1}$ values of $^{23}$Na known from the data of Ref.~%
\onlinecite{EPL2008} for the same phase of sodium cobaltates were long
enough to ascertain that they do not contribute significantly to the
transverse magnetization decay. So the values obtained by fitting the
relaxation decays for the sodium sites given in Table~\ref{tab:NaRelax} can
be associated with spin-spin processes.

The values of magnetization at zero time M(0) after $\nu ^{2}$ frequency
correction allowed us to estimate the relative intensities of sodium lines
(sites) which are also reported in Table~\ref{tab:NaRelax}. These data are
in good agreement with the more accurate ones obtained by NMR. Here the
accuracy \cite{EPL2009} is indeed limited primarily by the extrapolation
required from $\tau _{D}$ to $\tau=0$, and also by the variation of
spectrometer sensitivity with frequency.

\begin{table}[bp]
\caption{$^{23}$Na relaxation and NQR linewidth parameters.}%
\begin{ruledtabular}
\begin{tabular}{lcccc}
Site  & $I$~(\%) & $T_{2}$~($\mu $s) & $\Delta \nu_{calc}$~(kHz) & $\Delta \nu$~(kHz) \\
\hline
Na1  & 23(5)& 125(2)  & 2.5(1)  & 8.2(1)\\
Na2a & 46(8)& 82.3(7) & 3.8(2) & 8.21(5)\\
Na2b & 31(7)& 97(1)   & 3.3(2) & 9.02(6)\\
\end{tabular}
\end{ruledtabular}
\label{tab:NaRelax}
\end{table}

Due to the NSSR process the linewidth of an NQR signal at half-height should be equal $\Delta \nu_{calc}=1/(\pi \cdot T_{2})$ \cite{Abragam}. Such $\Delta \nu_{calc}$ for different Na sites are also collected in Table~\ref{tab:NaRelax}. And it is easy to see that they are approximately three times smaller than the experimentally measured linewidths for all sodium sites (Table~\ref{tab:NaRelax}). This clearly shows that the experimentally observable Na lines are inhomogeneously broadened by a weak distribution of EFG such that $\Delta \nu _{Q}/\nu _{Q}\approx 4\cdot 10^{-3}$. Such a small value found indeed imply a very well ordered structure in the sodium layers of this Na$_{2/3}$CoO$_{2}$ compound.

\subsubsection{$^{59}$Co spin-spin relaxation}

\begin{figure}[tbp]
\includegraphics[width=1.0\linewidth]{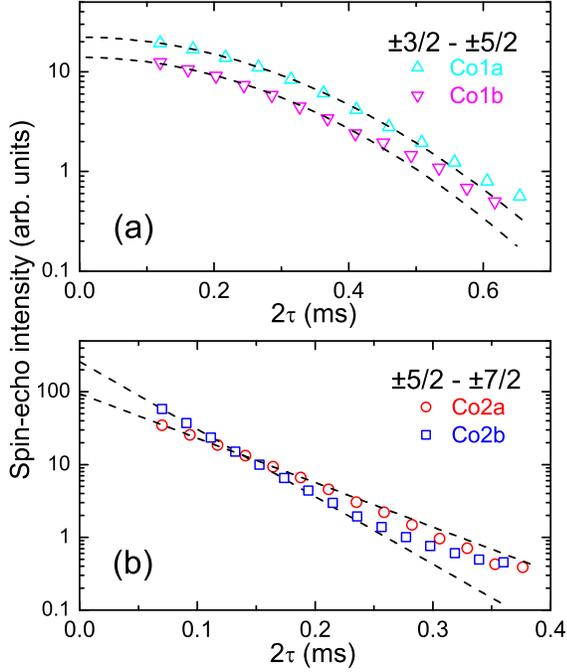}
\caption{Spin-spin relaxation curves for Co1a, b sites (a) on $\pm
3/2\leftrightarrow \pm 5/2$ transition, and for Co2a, b (b) on $\pm
5/2\leftrightarrow \pm 7/2$ transition.}
\label{fig:CoT2}
\end{figure}

The transverse relaxation decays for $^{59}Co$ sites are represented on
Figure~\ref{fig:CoT2}, where panel (a) shows data for the $\pm \frac{3}{2}%
\leftrightarrow \pm \frac{5}{2}$ transitions of the Co1a,b sites and panel
(b) that for the $\pm 7/2\leftrightarrow \pm 5/2$ transitions of the Co2a,b
sites. The results of the fit of these curves by Eq.~\ref{eq:funcT2} are
summarized in the Table~\ref{tab:CoRelax}. For the Co1 sites the transverse
magnetization decay has a Gaussian form ($n=2$), for Co2 it is a Lorentzian
with $n=1$. To understand this difference one needs to analyze the possible
influence of spin-lattice relaxation on the transverse relaxation.

For the cobalts the spin-lattice relaxation is fast enough so its
contribution to the spin-spin relaxation becomes important. As the
spin-lattice relaxation of $^{59}$Co is multi-exponential (Eq.~\ref%
{eq:Recovery}) then a good approximation might be done by considering that $%
T_{2}$ in the Eq.~\ref{eq:funcT2} should be related to the slope of the NSLR
curves $T_{1}^{\prime }$ at short times. For example, the Eq.~\ref%
{eq:Recovery} with the parameters from the Table~\ref{tab:CoT1Coef} can be
written, at short enough times for the Co2b's $\pm \frac{5}{2}%
\leftrightarrow \pm \frac{7}{2}$transition as
{\small
\[
R(t)\approx 1-\frac{10.12\,t}{T_{1}}=1-\frac{t}{T_{1}^{\prime}}.
\]%
}
In the Table~\ref{tab:CoRelax} one can see the estimated values of $%
T_{1}^{\prime }$ for different Co sites and transitions. For Co1a and Co1b
these $T_{1}^{\prime}$ times are much longer than the spin-spin relaxation $%
T_{2}$ on Fig.~\ref{fig:CoT2}a. Therefore the NSLR contribution to the NSSR
is negligible and we see in experiments the real spin-spin relaxation of Co1
sites. But for Co2, the calculated $T_{1}^{\prime }$ values become
comparable with the obtained $T_{2}$ times. To emphasize this we reported in
the Table~\ref{tab:CoRelax} the ratio $\alpha =T_{2}/T_{1}^{\prime }$. As
one can see $\alpha $ is almost the same for the two measured transitions $%
\pm \frac{5}{2}\leftrightarrow \pm \frac{3}{2}$ and $\pm \frac{7}{2}%
\leftrightarrow \pm \frac{5}{2}$ \ for the Co2b site. So we conclude that
for Co2 sites\ the measured $T_{2}$ values are determined by spin-lattice
relaxation process. This explains as well the difference in the transverse
magnetization decay shape for Co1 and Co2 sites.

This influence of spin-lattice relaxation on the spin-spin relaxation
explains why the intensity of the low frequency transitions of Co2a and Co2b
sites are found so weak experimentally (see Fig.~\ref{fig:CoSpec}). As one
can see from the Table~\ref{tab:CoRelax}, for the Co2b site, the measured $%
T_{2}=$30(2)~$\mu s$ for the $\pm \frac{3}{2}\leftrightarrow \pm \frac{5}{2%
}$ transition is \ smaller than that (49(2)~$\mu s$) for the $\pm \frac{5}{2%
}\leftrightarrow \pm \frac{7}{2}$ transition. Also as was already stated in
section III, the minimum usable time $\tau _{D}$ between RF pulses increases
with decreasing frequency. So both factors reduce the intensities of the
observed low frequency Co2 NQR lines.

As has been noticed previously in section III, the signal intensities
observed at different frequencies are affected by the variation of
spectrometer sensitivity, so careful determination of the relative site
occupancies of Co1 and Co2 could not be done directly from our experiments.
But we could estimate the fractional occupancies inside each group of cobalt
(Co1 and Co2) separately, as their resonance frequencies are close. After
correcting for the spin-spin relaxation $T_{2}$ and for the $\nu ^{2}$
frequency dependence of the signal intensity we obtained:
Co1a/Co1b=1.95(0.1) and Co2b/Co2a=1.9(0.2), in agreement with similar
determinations from $^{59}$Co NMR, which allowed as well better
determinations of the Co1/Co2 ratio \cite{EPL2008}.

Spectral lines of cobalt are inhomogeneously broadened (see Table~\ref%
{tab:CoRelax}) and this is clear if one compares the experimental linewidth
with the calculated one. Notice that the ratio $\Delta \nu /\Delta
\nu_{calc}$ is $\approx $ 4.0(0.4) for both $\pm \frac{3}{2}%
\leftrightarrow \pm \frac{5}{2}$ and $\pm \frac{5}{2}\leftrightarrow
\pm \frac{7}{2}$) transitions of Co2b and $\approx $ 5.0(0.3) for
Co2a, which confirms that these broadenings have the same origin.

\section{Conclusion}

In summary, we have reported detailed powder X ray and NQR data for the Na$%
_{2/3}$CoO$_{2}$ compound supporting the recent paper \cite{EPL2009} in
which we proposed a detailed structural model for this material from NMR/NQR
results. The NQR spectrum of $^{23}$Na and $^{59}$Co consists of several
narrow lines, which could be associated with 3 sodium and 4 cobalt sites, in
a 2D unit cell comprising 8 Na sites over 12 Co. The finite number of
unequivalent site positions confirms that the simple 3D ordering of the Na
layers leads to a differentiation of cobalt sites. We have performed as well
a careful Rietveld analysis of the X ray data which confirms the atomic 3D
unit cell proposed from NMR/NQR. The best Rietveld fits allowed us to
determine the slight atomic displacements of the Na atoms in the model unit
cell. Also the quality of the fit allows us to ascertain the x=2/3 value of
the Na content.

Others have also detected this phase and identified it with the
present one from its large spin susceptibility at low temperatures \cite{ChouPRL}. They have performed single crystal diffraction experiments in which they also find a 12 Co 2D unit cell. Using chemical analyzes of their single crystals they proposed that $x=0.71$, and have elaborated then a structural model in which the concentration of Na alternates between planes with $x=2/3$ with 8 Na per 2D cell and $x=3/4$, with 9 Na per 2D cell. In these Na layers the Na vacancies would be alternatively organized in divacancies and
trivacancies. One can immediately see that such a structure would never
explain the present NMR/NQR data as it corresponds to a larger number of
distinct Na sites (3 in each Na plane) than found experimentally. Furthermore the only sites exhibiting the threefold symmetry and $\eta =0$ would then be the Na1 sites of the $x=2/3$ planes. The fraction of such sites would be $x=2/17$, nearly twice smaller than the $25\%=2/8$ found experimentally.

The spin-spin and spin-lattice relaxation of the NQR lines were studied for
all Co sites, and the data analysis has allowed us to demonstrate that both $%
T_{1}$ and $T_{2}$ data do allow to establish the difference of magnetic
properties of the two types of Co sites. As indicated in Ref.~%
\onlinecite{EPL2009}, they constitute two sublattices: a Kagome structure
for the magnetic sites and the complementary triangular lattice of non-magnetic Co sites involving 25\% of the Co.

Beyond the simple results obtained for this Na$_{2/3}$CoO$_{2}$ compound, the approach developed here will be certainly useful to perform structural determinations for pure phases with different Na contents. We have indeed isolated some of them, which display quite distinct ground state physical properties \cite{EPL2008}.

\section{Acknowledgments}

We thank N.~Blanchard for her help in synthesizing materials
and Y.~S.~Meng for numeourous exchange about their computations. We acknowledge financial support by the ANR (NT05-441913) in France and by the grant RNP-6183 in Russia. Expenses in Orsay for I.R.M. and A.V.D. have been supported by the
"Triangle de la Physique". T.A.P. has obtained a fellowship
from the E.U. Marie Curie program \textquotedblleft Emergentcondmatphys\textquotedblright for part of her PhD work performed in Orsay.

\end{document}